# ASSESSING THE ASSOCIATION BETWEEN TRENDS IN A BIOMARKER AND RISK OF EVENT WITH AN APPLICATION IN PEDIATRIC HIV/AIDS

By Elizabeth R. Brown[1]

*University of Washington*

We present a new joint longitudinal and survival model aimed at estimating the association between the risk of an event and the change in and history of a biomarker that is repeatedly measured over time. We use cubic B-splines models for the longitudinal component that lend themselves to straight-forward formulations of the slope and integral of the trajectory of the biomarker. The model is applied to data collected in a long term follow-up study of HIV infected infants in Uganda. Estimation is carried out using MCMC methods. We also explore using the deviance information criteria, the conditional predictive ordinate and ROC curves for model selection and evaluation.

**1. Introduction.** In longitudinal studies it is common to monitor one or more biomarkers repeatedly over time while following participants until the occurrence of an event. Researchers are often interested in examining both the repeated measures and the time to event to gain an understanding of the underlying disease process. Additionally, the risk of an event may not depend solely on the level of the biomarker but also on the rate at which that biomarker is changing or its past average level. For example, two patients may present with the same biomarker value, but one patient's biomarker trajectory may be increasing while the other's is remaining constant. Their prognosis may appear to be be the same if only the current value is accounted for, but in fact the patient with the increasing biomarker may be at higher risk. In this paper we present a model to estimate the association between the risk of an event and the current value, as well as the rate of change or history of a longitudinally sampled biomarker. We illustrate the approach from a sub-study of HIVNET 012 [Guay et al. (1999); Jackson et al. (2003)]

Received June 2008; revised April 2009.
[1]Supported by a Contract from the National Institutes of Health NIAID U01 AI46702.

of HIV disease progression in Ugandan children who acquired HIV vertically, either in utero, during delivery or via breastmilk. In this sub-study children who tested positive before 18 months of age were followed until five years of age with blood samples drawn every 6 months. From these samples, the lab determined viral load, total lymphocyte count (TLC), CD4 percent and other related biomarkers. One aim was to determine the predictive value of these measures for time until death. In this manuscript we explore the association between change in and history of these markers and the risk of death. Overall, 128 children were followed after their first positive HIV test with lab measurements taken at regular intervals. Of these 128 infants, 70 died during follow-up.

In estimating the association between trends in a biomarker or set of biomarkers and the risk of an event, we face two important and distinct challenges. The first is selecting the correct model when the biomarker is collected in discrete time with error. The second is to determine how to obtain different summaries of the biomarker(s) and include them in the time to event model.

The first issue and its resolution through joint modeling of the time to event and longitudinal marker has been studied extensively as summarized by Tsiatis and Davidian (2004) when dealing with the current level of a biomarker and the risk of an event. In summary, survival analyses with time-dependent covariates can be biased if we simply include the raw measurements in the survival analysis [Prentice (1982)]. Joint longitudinal and survival models resolve this issue by modeling the biomarker process over time and including subject-specific parameters from the longitudinal model as covariates in the survival model. These same issues exist when trying to include other summaries of the biomarker process in the survival model, and may, in fact, be exacerbated.

Wulfsohn and Tsiatis (1997) and Faucett and Thomas (1996) introduced likelihood-based methods for analyzing a longitudinal marker and its association with the time to event simultaneously. This class of models is not restricted to linking the time-varying value of the longitudinal marker to the time-varying hazard through a regression on the current value of longitudinal model. They may instead include other information from the longitudinal trajectories summarized by the longitudinal model. Several authors have proposed models that group the trajectories into latent classes, then link the hazard and the longitudinal model by the latent class [Lin et al. (2002); Proust-Lima, Letenneur and Jacqmin-Gadda (2007); Han, Slate and Pena (2007); Proust-Lima et al. (2009)]. These models can help characterize longitudinal trajectories that may indicate a higher risk for event. In these models the latent class for the trajectory of the biomarker is defined based on the entire follow-up for the biomarker, which may have the drawback of using information about the biomarker from the future to estimate risk in the present.

Other descriptors from the longitudinal model have also been used to link the trajectory of the biomarker to risk of event. Yu, Taylor and Sandler (2008) developed individual risk prediction models for prostate cancer recurrence based on PSA trajectories. PSA was modeled using a nonlinear exponential decay and exponential growth model. The current value as well as slope (first derivative) of the PSA trajectory were included as time-varying covariates in the hazard model. Ye, Lin and Taylor (2008) presented a two-stage regression calibration approach for modeling longitudinal and time-to-event data. Their longitudinal model included smoothing splines at the population level with individual deviations from the smoothing spline allowed through a mean 0 integrated Wiener process and subject-specific slopes and intercepts. Both the current value and slope of the subject-specific trajectories were included as covariates in the hazard model.

In this manuscript we extend the work of Brown, Ibrahim and DeGruttola (2005) who used cubic B-splines to model the impact of multiple biomarkers on time to event to include the slope and integral of the cubic B-spline models as time-varying covariates in the hazard model. They showed that cubic B-splines provided flexibility in the biomarker model that simple parametric models could not. Because cubic B-spline trajectory models and their slopes and integrals are linear functions of the parameters that do not increase exponentially with time, they avoid computational instability. Additionally, because the basis functions of the cubic B-splines weight more heavily on local (in time) information for estimating the value of the trajectory, estimates of the slope early in follow-up do not rely heavily on values of the biomarker observed late in follow-up.

The paper proceeds as follows. In the next section we review cubic B-spline models for longitudinal data and outline the model associating change and cumulative exposure with time to event. Next, we discuss estimation and model selection procedures. We then show an example from HIVNET 012. We conclude with a discussion.

**2. The joint longitudinal and survival model.** In this section we describe a joint longitudinal and survival model to estimate the association between the rate of change in a biomarker or cumulative history of a biomarker and the risk of an event. We begin with a description of the notation and a review of the longitudinal cubic B-spline model, including expressions for the first derivatives and integrals. We then introduce the model linking the biomarker and its rate of change or cumulative history to the risk of event.

2.1. *The longitudinal model.* Let $y_{ijl}$ denote the $i$th, $i = 1, \ldots, N$, subject's $j$th, $j = 1, \ldots, m_i$, observation of the $l$th, $l = 1, \ldots, L$, biomarker at time $t_{ij} < T$, where $T$ denotes the end of follow-up. We define an observation at time $t_{ij}$ to be a function of the true underlying trajectory $\psi(t_{ij})$ plus error,

$$y_{ijl} = \psi_l(t_{ij}) + e_{ijl},$$

where the errors are independent and normally distributed such that $(e_{ij1}, \ldots, e_{ijL})' \sim N_l(0, \Sigma)$, where $N_l(a, b)$ is the $l$-dimensional multivariate normal distribution with mean vector $a$ and covariance matrix $b$. Brown, Ibrahim and DeGruttola (2005) modeled the true, but unobserved, trajectory using cubic B-splines such that

$$\psi_l(t) = \sum_{k=1}^{q} \beta_{ilk} B_k(t), \tag{2.1}$$

where $\{B_k(\cdot)\}$ is a $\bar{q}$-dimensional basis for spline functions on $[0, T]$ with a fixed knot sequence, $u = (u_1, \ldots, u_{q+4})$ and $\beta_{il} = (\beta_{il1}, \ldots, \beta_{ilq})'$ is a vector of subject-specific parameters of length $q$ that determine the shape of the $i$th subject's trajectory. We assume a hierarchical model where $\beta_{il} \sim N_l(b_{0l} + X_i'\alpha_l, V_{0l})$. In this model the effect of the covariates is modeled at the population level where $\alpha_l$ is a vector of parameters of length $p$ linking the vector of baseline covariates $X_i$ to the longitudinal outcome and $b_0$ is the vector of length $q$ of the mean of the coefficients for the $k$th basis function when $X_{i1}, \ldots, X_{ip} = 0$. Brown, Ibrahim and DeGruttola (2005) showed the merits (better mixing in the Gibbs sampler and more flexible effect of the covariate on the longitudinal outcome) of modeling the effect of covariates on the longitudinal model in this level of the hierarchy. As a further extension, we allow for a covariance structure for the spline coefficients where $V_{0l}$ is the $q \times q$ covariance matrix of the coefficients of the basis functions.

An individual's contribution to the likelihood of the longitudinal marker can be expressed as

$$p(Y_i | \Sigma, \beta_i) \propto \frac{1}{|\Sigma|^{m_i}} \exp\left\{-\frac{1}{2} \sum_{j=1}^{m_i} (Y_{ij} - \psi(t_{ij}))' \Sigma^{-1} (Y_{ij} - \psi(t_{ij}))\right\}, \tag{2.2}$$

where $Y_{ij} = (y_{ij1}, \ldots, y_{ijL})'$ and $\psi(t_{ij}) = (\psi_1(t_{ij}), \ldots, \psi_L(t_{ij}))'$.

2.2. *The joint model.* We next review the form of the joint model when we are only interested in modeling the relationship between the level of the marker at time $t$ and the hazard at time $t$. We then show how the spline model of the trajectory can be used to describe the rate of change in and history of the biomarker. Next, we incorporate these functionals into the hazard, thus extending the joint model to estimate the impact of the rate of change and history on the risk of an event.

The usual joint longitudinal and survival model assumes proportional hazards, and the effect of the trajectories of the biomarkers on the hazard is modeled as

$$\lambda(t) = \lambda_0(t) \exp(\gamma' \psi(t) + Z_i' \zeta), \tag{2.3}$$

where $\lambda_0(t)$ is the baseline hazard at time $t$, $\gamma$ is a parameter vector of length $L$ linking the trajectory vector at time $t$ to the hazard at time $t$, and $\zeta$ is a vector of parameters of length $p_z$ linking the vector of baseline covariates $Z_i$ to the hazard. Taking a likelihood approach requires some specification of the baseline hazard. Here, we specify a piecewise constant hazard allowing for approximately 8–10 events in each interval, where

$$\lambda_0(t) = \lambda_j, \qquad w_j \le t < w_{j+1}, j = 1, \dots, J,$$

where the $w_j$'s are the jump points with $w_1 = 0$ and $w_{J+1} = \infty$.

Then the cumulative hazard,

$$\int_0^{s_i} \lambda(u) e^{\gamma' \psi(u) + z_i' \zeta}\, du,$$

can be rewritten as

$$e^{z_i' \zeta} \sum_{j=1}^{J} H_{ij}(\beta, \gamma, \lambda),$$

where

$$H_{ij}(\beta, \gamma, \lambda) = I\{s_i \ge u_{j-1}\} \lambda_j \int_{u_{j-1}}^{\min(u_j, s_i)} e^{\gamma' \psi(u) + z_i' \zeta}\, du \tag{2.4}$$

and $I\{s_i \ge u_{j-1}\}$ is an indicator function which equals 1 if the event time occurs in or later than the $j$th interval and 0 otherwise. The integral in (2.4) does not have an analytical solution for the trajectory defined by cubic B-splines. Instead, for computational ease and speed, we use Gaussian quadrature to approximate it.

Extending the model to include the relationship between another function or functions of the time-varying covariates and the hazard requires adding another term to the model. If we are interested in the relationship between rate of change of the biomarker and the hazard, we include the first derivative of $\psi(t)$. If we are interested in the relationship between the cumulative history of the biomarker and the hazard, we include the integral of $\psi(t)$.

For clarity, we drop the subscripts for definition of the first derivative and integral. The first derivative of the trajectory function as shown by de Boor [(2001), page 116] can be expressed as

$$\psi'(t) = B^{(2)*}(t)' \beta, \tag{2.5}$$

where $B^{(2)*}(t)$ is a vector of length $q$ with the $j$th element equal to $\frac{B_j^{(2)}(t)}{u_{j+3}-u_j} - \frac{B_{j+1}^{(2)}(t)}{u_{j+4}-u_{j+1}}$, and $B^{(2)}(t)$ is the quadratic B-spline based on the same sequence of knots as the original B-spline in (2.1) with the first and last knot removed. Equation (2.5) is written as a linear function of the elements of $\beta_i$; however, it is a quadratic B-spline and still retains many of the desirable properties of B-splines mentioned earlier.

The general idea for calculating the integral of the cubic B-spline was laid out by de Boor [(2001), page 128]. We derived the integral of the trajectory up to time $t$ to be

$$(2.6) \qquad \int_0^t \psi(v)\,dv = \psi^{(-1)}(t) = B^{*(4)}(t)'\beta,$$

where $B^{*(4)}(t)$ is a vector of length $q$ with $j$th element equal to $\sum_{k=j+1}^{q+1} B_k^{(4)}(t) \times (\frac{u_{j+4}-u_j}{4})$, and $\{B_k^{(4)}(\cdot)\}$ is the vector of $q+1$-dimensional basis of a quartic B-spline based on the knot vector $u$ augmented by two arbitrary knots, $u_0 < u_1$ and $u_{q+5} > u_{q+4}$.

To link the rate of change and history of the trajectories to the hazard, we use the following regression model:

$$(2.7) \qquad \lambda(t) = \lambda_0(t) \exp\{\gamma'\psi(t) + \gamma_s'\psi'(t) + \gamma_h'\psi^{(-1)}(t) + Z_i'\zeta\},$$

where $\gamma_s$ is the $L$-length vector of parameters linking the $L$-length vector of the slopes of the trajectories at time $t$, $\psi'(t)$ to the hazard at time $t$ and $\gamma_h$ is the $L$-length vector of parameters linking the $L$-length vector of the cumulative histories of the trajectories up to time $t$, $\psi^{-1}(t)$ to the hazard at time $t$.

As in the case where only the trajectory value at time $t$ is linked to the hazard at time $t$, the cumulative hazard can be written as

$$e^{z_i'\zeta} \sum_{j=1}^{J} H_{ij}(\beta, \gamma, \gamma_s, \gamma_h, \lambda),$$

where

$$(2.8) \qquad \begin{aligned} &H_{ij}(\beta, \gamma, \gamma_s, \gamma_h, \lambda) \\ &= I\{s_i \geq u_{j-1}\}\lambda_j \\ &\quad \times \int_{u_{j-1}}^{\min(u_j, s_i)} \exp\{\gamma'\psi(u) + \gamma_s'\psi'(u) + \gamma_h'\psi^{(-1)}(u)\}\,du \end{aligned}$$

and $\gamma_s = (\gamma_{s1}, \ldots, \gamma_{sL})'$ and $\gamma_h = (\gamma_{h1}, \ldots, \gamma_{hL})'$. Because equation (2.8) does not have a closed form solution, we evaluate it numerically using Gaussian quadrature.

The distribution for an individual's time to event, $s_i$, given the trajectory function and choice of hazard regression, is given by

$$(2.9) \quad f(s_i, \nu_i | Y_i) = \lambda(s_i)^{\nu_i} \exp\{\nu_i(\gamma' \psi(s_i) + \gamma'_s \psi'(s_i) + \gamma'_h \psi^{(-1)}(s_i) + z'_i \zeta)\}$$
$$\times \exp\left\{-e^{z'_i \zeta} \sum_{j=1}^{J} H_{ij}(\beta, \gamma, \gamma_s, \gamma_h, \lambda)\right\},$$

where $\nu_i$ is the censoring indicator for subject $i$.

**3. Estimation.** We can now express the $i$th subject's contribution to the joint likelihood function as

$$f(Y_i, s_i, \nu_i) = f(s_i, \nu_i | Y_i) \times f(Y_i),$$
$$f(Y_i, s_i, \nu_i) \propto \lambda(s_i)^{\nu_i} \exp\{\nu_i(\gamma' \psi(s_i) + \gamma'_s \psi'(s_i) + \gamma'_h \psi^{(-1)}(s_i) + z'_i \zeta)\}$$
$$\times \exp\left\{-e^{z'_i \zeta} \sum_{j=1}^{J} H_{ij}(\beta, \gamma, \gamma_s, \gamma_h, \lambda)\right\}$$
$$\times \frac{1}{|\Sigma|^{m_i}} \exp\left\{-\frac{1}{2} \sum_{j=1}^{m_i} (Y_{ij} - \psi(t_{ij}))' \Sigma^{-1} (Y_{ij} - \psi(t_{ij}))\right\}.$$

We then specify the prior distributions for the parameters in the likelihood as follows. We assume $(\gamma, \gamma_s, \gamma_h)' \sim N_{3L}(G_0, G_1)$, $\Sigma \sim \text{Wishart}_{\nu_\Sigma}(S_\Sigma^{-1})$ and $\lambda_j \sim \text{gamma}(d_{0j}, d_{1j})$. We may also specify priors on the hyperparameters of $\beta$, $\alpha_l \sim N_p(C_{0l}, C_{1l})$, $b_{0l} \sim N_p(A_{0l}, A_{1l})$ and $V_{0l}^{-1} \sim \text{Wishart}_{\nu_{v_{0l}}}(S_{v_{0l}}^{-1})$. Here, $\text{Wishart}_\nu(S^{-1})$ denotes the Wishart distribution with $\nu$ degrees of freedom and scale matrix $S^{-1}$ and $\text{gamma}(a, b)$ denotes the gamma distribution with shape parameter $a$ and scale parameter $b$. The prior distributions were chosen to be as general as possible while still being proper and conjugate to the likelihood when possible.

We use the Gibbs sampler [Gelfand and Smith (1990)] to sample from the posterior distribution of the parameters. Because the full conditionals of $\gamma$, $\gamma_s$, $\gamma_h$ and $\zeta$ are log-concave, we use adaptive rejection sampling (ARS) [Gilks and Wild (1992)] to sample from them. We use the slice sampler [Neal (2003)] to sample the random effects, $\beta_{ij}, j = 1, \ldots, q, i = 1, \ldots, N$. The full conditionals of $\lambda$, $\Sigma$, $V$ and $b$ are sampled from directly. The estimation procedure is implemented in R [R Development Core Team (2006)] and C with code available from the author.

**4. Model comparison.** We examine two statistics for model comparison, the deviance information criterion (DIC) [Spiegelhalter et al. (2002)] and the conditional predictive ordinate (CPO) [Gelfand, Dey and Chang (1992)].

The DIC is a measure of the deviance penalized for the number of parameters in the model which may be difficult to ascertain in hierarchical models and is therefore estimated. The DIC is the sum of the deviance estimated using the posterior estimates of the parameters, $D(\bar{\Theta})$, and twice the effective number of parameters, $p_D$. The effective number of parameters is estimated by $p_D = \overline{D(\Theta)} - D(\bar{\Theta})$, where $\overline{D(\Theta)}$ is the posterior mean of the deviance (the average of the deviances calculated using the estimated parameters at each step of the MCMC sampler). For the model presented in this paper, the DIC can be expressed as

$$DIC = 2\frac{1}{G}\sum_{g=1}^{G}\sum_{i=1}^{N}\log\{f(s_i,\nu_i,Y_i|\Theta^{(g)})\} - \sum_{i=1}^{N}\log\{f(s_i,\nu_i,Y_i|\bar{\Theta})\},$$

where $\Theta^{(g)}$ denotes the parameter samples at the $g$th, $g=1,\ldots,G$, iteration of the Gibbs sampler and $\bar{\Theta}$ represents the means of the posterior samples. A smaller DIC indicates a better fit when comparing models.

For the $i$th observation, the CPO statistic is defined as

$$(4.1) \quad CPO_i = f(s_i,\nu_i,Y_i|D^{(-i)}) = \int f(s_i,\nu_i,Y_i|\Theta,D_i)\pi(\Theta|D^{(-i)})\,d\Theta,$$

where $\Theta$ denotes the model parameters, $D_i$ denotes the $i$th patient's covariate data and $D^{(-i)}$ denotes the covariate data for all the patients except the $i$th patient. We cannot obtain a closed form solution for (4.1); however, Chen, Shao and Ibrahim [(2000), Chapter 10] show that a Monte Carlo approximation of (4.1) is

$$\widehat{CPO_i} = \left(\frac{1}{G}\sum_{g=1}^{G}\frac{1}{f(s_i,\nu_i,Y_i|\Theta^{(g)})}\right)^{-1},$$

where $\Theta^{(g)}$ denotes the parameter samples at the $g$th, $g=1,\ldots,G$, iteration of the Gibbs sampler. A large $CPO$ value indicates a better fit. We can compare different models using the sums of the logs of the CPOs of the individual observations, also known as the log pseudo-marginal likelihood (LPML). Models with greater LPML $= \sum \log(\widehat{CPO_i})$ values will indicate a better fit. Both the DIC and LPML are designed to measure a model's predictive ability, although the DIC is based on a penalized deviance approach and the LPML is based on a cross-validated approach.

**5. Application.** HIVNET 012, conducted in Uganda, was a double-blinded controlled randomized trial of single dose nevirapine for the mother and newborn infant versus AZT administered only to the mother to prevent mother to child transmission (MTCT) of HIV. In spite of the success of nevaripine in reducing the risk of transmission, many infants still experienced MTCT of HIV. To better understand HIV infection in young children, 128 of those infants were enrolled in a long term follow-up study and followed until age 5 or death. CD4 cell percent and HIV viral load are known indicators of disease progression. TLC is also being studied for its potential use in resource-poor settings where routine CD4 and viral load monitoring may be cost prohibitive. In this section we examine the association between longitudinal measures of CD4 cell percent, viral load and TLC and time until death using separate models (one biomarker per model) in the HIVNET 012 long term follow-up study using the methods proposed in the previous sections.

Seventy infants died during follow-up. Jump points for the baseline hazard function were selected based on quantiles of the event times so that approximately 8 events occurred between the jump points. Infants had between one and thirteen longitudinal measures with a median of four. Overall, there were a total of 594 measurements of CD4 percent, 603 measurements of viral load and 763 measurements of TLC. 13% of the CD4 percent measures, 7% of the viral load measures and 16% of the TLC measures are taken at time 0. In this analysis we placed the knots for the splines based on the quantiles of the data; therefore, this clumping of measurement times limits the number of knots we could select before we start placing multiple knots at 0, making the slope undefined. Therefore, for CD4 percent, we fit models with $q = 5, \ldots, 10$. For TLC, we fit models with $q = 5, \ldots, 9$. For comparison to potential models with more basis functions, we also fit models with equally spaced knots with $q = 11$ for CD4 percent and $q = 10$ for TLC. For viral load, we can fit models with $q = 5, \ldots, 19$. Additionally, we included age at first positive HIV test as a covariate in the hazard model; therefore, the interpretation of $\zeta$ is the change in the log hazard associated with a 1 month increase in the age of the infant at the time of the first positive HIV test. We implemented the Gaussian quadrature procedure using 10 points. We also ran models with higher $q$ using 50 points and found no difference in the estimates.

Table 1 shows the estimated values of the parameters of interest along with their 95% credible intervals obtained from fitting the data to three versions of the hazard model shown in equation (2.7), one with $\gamma_s = 0$ and $\gamma_h = 0$, one with $\gamma_h = 0$ and one with $\gamma_s = 0$, for a range of $q$. The Gibbs sampler was run twice with different starting values and seeds for the random number generator for 100,000 iterations for each model with a burn-in of 10,000. Samples from every 10th iteration were saved to reduce possible autocorrelation. The resulting sample was used to compute parameter estimates and credible intervals as well as DIC and LPML. Based on DIC, the minimum number of basis functions for a cubic B-spline, $q = 5$, is not adequate for any of the three longitudinal outcomes. Although not shown, models with equally spaced knots never outperformed the models with knots based on quantiles according to DIC or LPML. There were no meaningful differences in the estimates between the results from the two samplers. Convergence was assessed using diagnostic tools provided in the CODA package [Plummer et al. (2006)].

TABLE 1
*Results from the models for the three markers of HIV disease progression*

|  |  | $\gamma$ | $\gamma_s$ | $\gamma_h$ ($\times 100$) | $\zeta$ | DIC | LPML |
|---|---|---|---|---|---|---|---|
| **VL** | | | | | | | |
| $q=5$ | current | 1.34 (0.87, 1.89) | | | −0.08 (−0.17, −0.01) | 1644 | −744 |
| | +slope | 1.57 (1.06, 2.18) | 5.23 (1.91, 8.62) | | −0.07 (−0.16, 0.01) | 1607 | −726 |
| | +history | 0.72 (−0.02, 1.43) | | 5.56 (1.08, 11.45) | −0.09 (−0.19, −0.02) | 1632 | −737 |
| $q=6$ | current | 1.45 (0.93, 2.03) | | | −0.08 (−0.18, −0.01) | 1635 | −741 |
| | +slope | 1.88 (1.32, 2.53) | 2.79 (1.48, 4.31) | | −0.07 (−0.17, 0.01) | 1590 | −717 |
| | +history | 0.95 (0.20, 1.73) | | 4.26 (−0.07, 9.89) | −0.09 (−0.20, −0.01) | 1633 | −736 |
| $q=7$ | current | 1.58 (1.06, 2.21) | | | −0.08 (−0.18, −0.01) | 1649 | −735 |
| | +slope | 2.03 (1.49, 2.70) | 1.49 (0.91, 2.25) | | −0.07 (−0.19, 0.02) | 1647 | −709 |
| | +history | 1.20 (0.47, 1.99) | | 3.28 (−0.97, 8.38) | −0.09 (−0.19, −0.01) | 1653 | −734 |
| $q=8$ | current | 1.64 (1.11, 2.28) | | | −0.09 (−0.18, −0.01) | 1646 | −731 |
| | +slope | 1.93 (1.34, 2.62) | 0.88 (0.29, 1.6) | | −0.09 (−0.19, 0) | 1616 | −715 |
| | +history | 1.22 (0.47, 2.04) | | 3.84 (−0.63, 9.59) | −0.09 (−0.21, −0.01) | 1655 | −728 |
| $q=9$ | current | 1.62 (1.09, 2.27) | | | −0.08 (−0.18, −0.01) | 1662 | −730 |
| | +slope | 1.92 (1.31, 2.66) | 0.98 (−0.32, 2.22) | | −0.08 (−0.19, 0.00) | 1633 | −714 |
| | +history | 1.26 (0.54, 2.07) | | 3.27 (−1.07, 8.78) | −0.10 (−0.20, −0.03) | 1668 | −725 |
| **CD4** | | | | | | | |
| $q=5$ | current | −0.83 (−1.17, −0.54) | | | −0.12 (−0.22, −0.05) | 1767 | −743 |
| | +slope | −0.87 (−1.22, −0.56) | −1.81 (−4.24, 0.37) | | −0.13 (−0.23, −0.05) | 1764 | −732 |
| | +history | −0.6 (−1.03, −0.19) | | −2.05 (−4.75, 0.36) | −0.13 (−0.23, −0.05) | 1747 | −731 |
| $q=6$ | current | −0.81 (−1.15, −0.51) | | | −0.12 (−0.22, −0.04) | 1738 | −741 |
| | +slope | −0.91 (−1.29, −0.58) | −1.47 (−3.32, 0.80) | | −0.12 (−0.22, −0.05) | 1728 | −730 |
| | +history | −0.57 (−1.02, −0.15) | | −1.76 (−4.14, 0.38) | −0.13 (−0.23, −0.05) | 1722 | −726 |
| $q=7$ | current | −0.96 (−1.34, −0.62) | | | −0.12 (−0.22, −0.05) | 1674 | −731 |
| | +slope | −0.98 (−1.39, −0.63) | −0.38 (−1.36, 0.82) | | −0.12 (−0.22, −0.05) | 1672 | −719 |
| | +history | −0.77 (−1.29, −0.29) | | −1.2 (−3.74, 1.03) | −0.13 (−0.23, −0.05) | 1666 | −723 |

TABLE 1
*(Continued)*

|  |  | $\gamma$ | $\gamma_s$ | $\gamma_h (\times 100)$ | $\zeta$ | DIC | LPML |
|---|---|---|---|---|---|---|---|
| $q=8$ | current | $-1.01\,(-1.42, -0.64)$ |  |  | $-0.13\,(-0.23, -0.05)$ | 1690 | $-732$ |
|  | +slope | $-1.03\,(-1.46, -0.66)$ | $-0.26\,(-1.09, 0.79)$ |  | $-0.13\,(-0.23, -0.05)$ | 1686 | $-719$ |
|  | +history | $-0.83\,(-1.4, -0.3)$ |  | $-1.14\,(-3.77, 1.23)$ | $-0.13\,(-0.23, -0.05)$ | 1684 | $-719$ |
| $q=9$ | current | $-0.94\,(-1.32, -0.59)$ |  |  | $-0.12\,(-0.22, -0.05)$ | 1678 | $-733$ |
|  | +slope | $-0.94\,(-1.34, -0.59)$ | $0.07\,(-0.54, 0.74)$ |  | $-0.12\,(-0.22, -0.05)$ | 1678 | $-721$ |
|  | +history | $-0.74\,(-1.25, -0.24)$ |  | $-1.95\,(-4.65, 0.47)$ | $-0.13\,(-0.23, -0.05)$ | 1666 | $-723$ |
| TLC |  |  |  |  |  |  |  |
| $q=5$ | current | $-0.20\,(-0.42, 0.01)$ |  |  | $-0.11\,(-0.19, -0.04)$ | 3653 | $-1387$ |
|  | +slope | $-0.31\,(-0.57, -0.09)$ | $-1.77\,(-3.28, -0.04)$ |  | $-0.11\,(-0.20, -0.04)$ | 3617 | $-1376$ |
|  | +history | $-0.22\,(-0.52, 0.06)$ |  | $0.12\,(-1.17, 1.32)$ | $-0.10\,(-0.19, -0.04)$ | 3649 | $-1380$ |
| $q=6$ | current | $-0.45\,(-0.72, -0.19)$ |  |  | $-0.11\,(-0.21, -0.05)$ | 3485 | $-1300$ |
|  | +slope | $-0.47\,(-0.75, -0.23)$ | $-0.34\,(-0.73, 0.04)$ |  | $-0.11\,(-0.20, -0.04)$ | 3474 | $-1288$ |
|  | +history | $-0.70\,(-1.04, -0.38)$ |  | $1.46\,(0.16, 2.78)$ | $-0.11\,(-0.19, -0.04)$ | 3458 | $-1288$ |
| $q=7$ | current | $-0.39\,(-0.64, -0.16)$ |  |  | $-0.11\,(-0.21, -0.04)$ | 3521 | $-1297$ |
|  | +slope | $-0.40\,(-0.65, -0.16)$ | $-0.11\,(-0.32, 0.11)$ |  | $-0.11\,(-0.19, -0.04)$ | 3516 | $-1287$ |
|  | +history | $-0.54\,(-0.86, -0.24)$ |  | $0.96\,(-0.30, 2.21)$ | $-0.11\,(-0.20, -0.04)$ | 3504 | $-1287$ |
| $q=8$ | current | $-0.51\,(-0.75, -0.28)$ |  |  | $-0.12\,(-0.21, -0.04)$ | 3465 | $-1261$ |
|  | +slope | $-0.35\,(-0.62, -0.10)$ | $0.02\,(-0.07, 0.14)$ |  | $-0.10\,(-0.19, -0.04)$ | 3451 | $-1265$ |
|  | +history | $-0.66\,(-0.97, -0.40)$ |  | $1.24\,(0.04, 2.48)$ | $-0.11\,(-0.19, -0.04)$ | 3451 | $-1248$ |
| $q=9$ | current | $-0.45\,(-0.68, -0.22)$ |  |  | $-0.10\,(-0.19, -0.04)$ | 3367 | $-1200$ |
|  | +slope | $-0.44\,(-0.68, -0.21)$ | $0.01\,(-0.04, 0.08)$ |  | $-0.10\,(-0.19, -0.04)$ | 3368 | $-1191$ |
|  | +history | $-0.54\,(-0.80, -0.27)$ |  | $0.77\,(-0.41, 1.90)$ | $-0.11\,(-0.19, -0.04)$ | 3367 | $-1192$ |

For viral load, DIC selected $q = 6$ for all three models. The LPML increased as $q$ increased except for the model with $\gamma$ and $\gamma_s$ where it selected $q = 7$; however, the value was not very different than that for $q = 6$. Here, we focus on the results for $q = 6$. The point estimate of $\gamma$ indicates an increased risk of death with increasing viral load such that a 10-fold increase in viral load is associated with a 16.3-fold increase in the hazard [95% credible interval (CI): (4.4, 74.5)]. The rate of change in viral load is also associated with risk of death. At a known level of viral load, a unit increase per month in the rate of change of log viral load is associated with a 2.8-fold increase in the hazard [95% CI: (1.15, 7.4)]. The estimate of the strength of this association decreases as $q$ increases. However, increasing $q$ also introduces more fluctuations in the estimate of the slope over time that may not be supported by the data. Additionally, the history of viral load is also associated with risk of death, with a one unit increase in the mean value of the trajectory of log viral load times a unit change in the length (in months) of follow-up being associated with 1.04-fold increase in the hazard. In plainer terms, if two infants have been followed for 7 months, with a difference in mean values equal to 1, the hazard ratio would be $\exp(7*4.3/100) = 1.35$. If after 12 months the difference in mean levels was still equal to 1, the hazard ratio would be 1.68. Figure 1 shows the trajectory fits for viral load with $q = 6$. The model fits a variety of shapes suggested by the data. For example, the initial steep rise in infants 2, 10 and 12 is fit well, as is the initial decrease in infants 6 and 9.

For CD4 percent, the DIC selected $q = 7$ for all three models. The LPML selected $q = 7$ for the current value and current value plus slope models and $q = 8$ for the current value plus history model (although the value was close to that for $q = 7$). Here, we focus on the results for $q = 7$. A 10 unit decrease in CD4 percent is associated with a 2.6-fold increase in the hazard [95% CI: (1.86, 3.82)]. There is no suggestion of a further association between risk of death and the change in CD4 percent or its history. Figure 2 shows the fit of the spline model to the observed CD4 percent data for 12 infants. The majority of the infants in Figure 2 have an initial sharp decline in CD4 percent which is fit well by the model without forcing decreases where the data do not suggest it (infant 6).

For TLC, the DIC and LPML selected $q = 9$. The density of measurements near time zero forced quantile-based knots very close together if not equal shortly after infection leading to undefined slopes for models with $q > 9$. Using equally spaced knots did not result in improved DIC or LPML. Here we focus on the results for $q = 9$. A possible association between TLC and risk of death was suggested in this model with a 1000 unit increase in TLC being associated with a 34% [95% CI: (20, 49)] decrease in the risk of death. Figure 3 shows the fit from the longitudinal models for TLC.

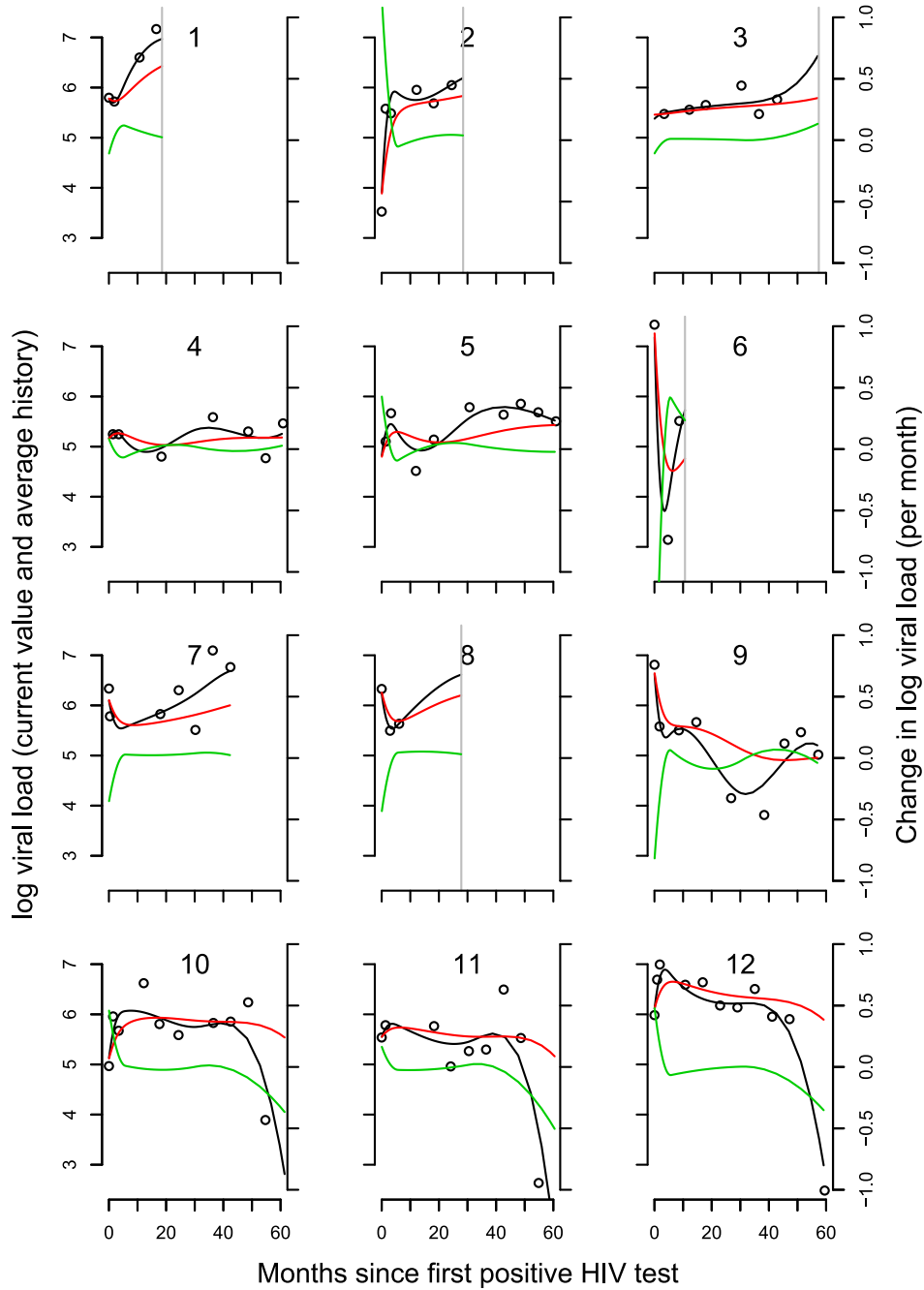

FIG. 1. *Longitudinal profiles of viral load for 12 infants. Circles mark the observed values. The solid black line indicates the fit from the model. The red line represents the fitted cumulative value. The green line represents the fitted slope. A vertical grey line indicates time of death if it was observed.*

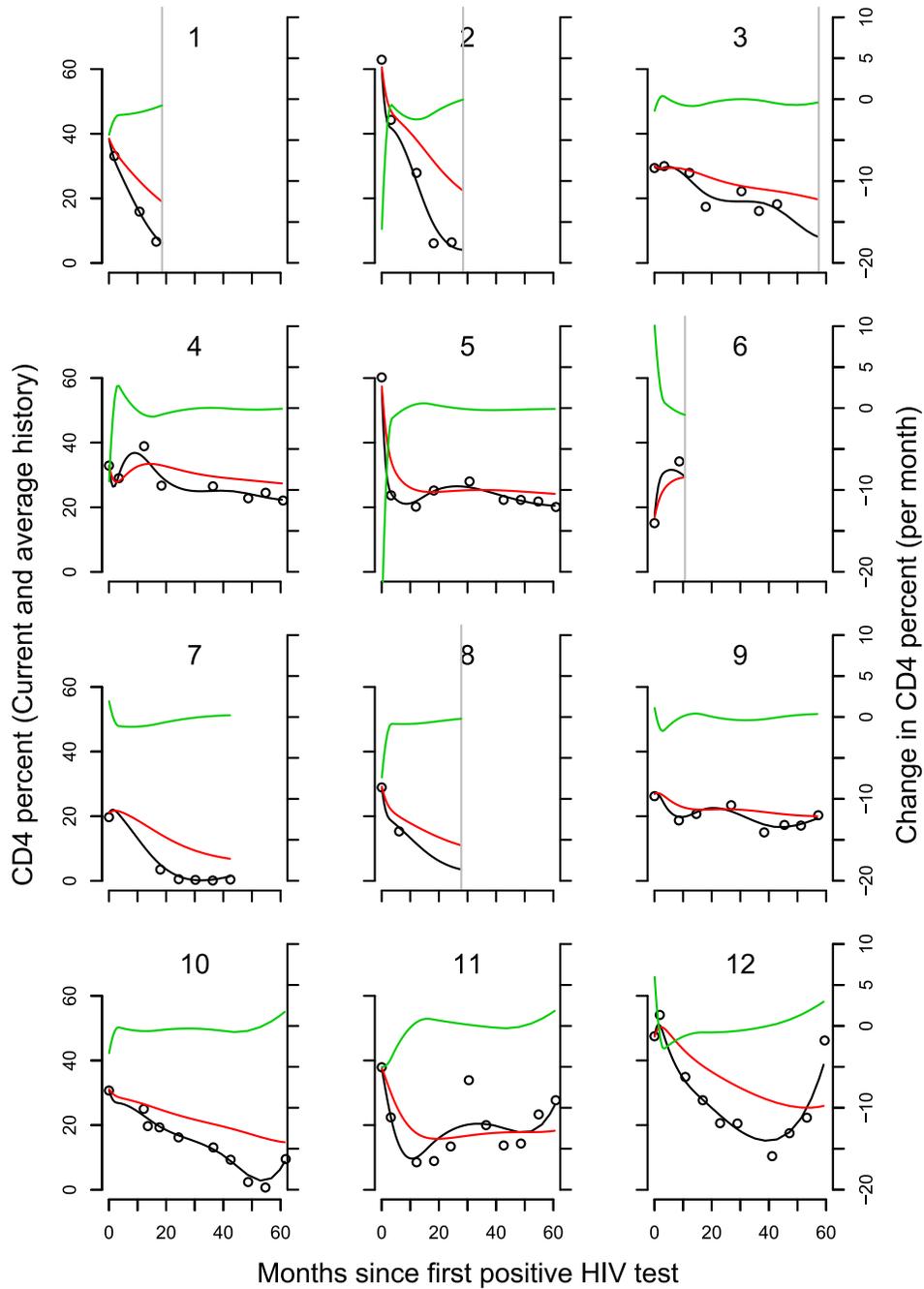

FIG. 2. *Longitudinal profiles of CD4 percent for 12 infants. Circles mark the observed values. The solid black line represents the fit from the model. The red line represents the fitted cumulative value. The green line represents the fitted slope. A vertical grey line indicates time of death if it was observed.*

The models all estimate a similar association between age at first positive test and risk of death, with a one month increase in age at first positive test associated with an approximate 10% reduction in risk given the trajectory of the longitudinal marker.

Figure 4 shows the posterior estimate of the cumulative hazards for the three markers when only the current value of the marker is included in the model. TLC provide better fits to the Kaplan–Meier estimate in the first year, while viral load provides a better fit between 2 and 3 years.

We propose using ROC curves for censored data [Heagerty, Lumley and Pepe (2000)] as an additional comparison of joint longitudinal and survival models. We plotted ROC curves for predicting death within 1 year based on linear predictor, $\gamma\psi(t) + \gamma_s\psi'(t) + \gamma_h\psi^{(-1)}(t)) + Z_i'\zeta$, at $t = 6$, 12 and 18 months after infection (Figure 5). Taking $t = 6$ months as an example, we treat the baseline time for survival as 6 months, and calculate the ROC curve for death within 12 months (18 months after infection) with the linear predictor calculated at 6 months as the biomarker. Infants who died or were censored before 6 months are not included. The same procedure was used for 12 and 18 months. These results do not suggest any large improvement in prediction when either the rate of change in or history of the biomarker are included in the models, except possibly for including slope in the viral load model. This agrees with the results from the model and the DIC and LPML. Overall, TLC does not compare favorably to either viral load or CD4 percent in predicting death within one year, and viral load appears to be the best predictor.

**6. Discussion.** Understanding the relationship between trends in a biomarker and risk of an event can yield important insight into the mechanisms of disease progression. Clearly, this is recognized scientifically, as the *Journal of the American Medical Association* recently made an exception to its own policies on publishing data over five years old to report an analysis of the association between CD4 slopes estimated from linear mixed effects models and time to development of AIDS or death in antiretroviral naive HIV-infected adults [Mellors et al. (2007)]. Here, we present a model that goes several steps further to addressing that question by jointly modeling the time to event and the slope and looking at local changes as opposed to long term trends. Additionally, we do not restrict ourselves to a linear model that assumes constant rate of change. This model is motivated and illustrated by a long-term follow-up study of HIV-infected children in Africa. However, it can be used in many settings where understanding the relationship between trends in a biomarker and time to an event is of interest. Other examples may include cognitive function and death or subclinical coronary disease and clinical coronary events.

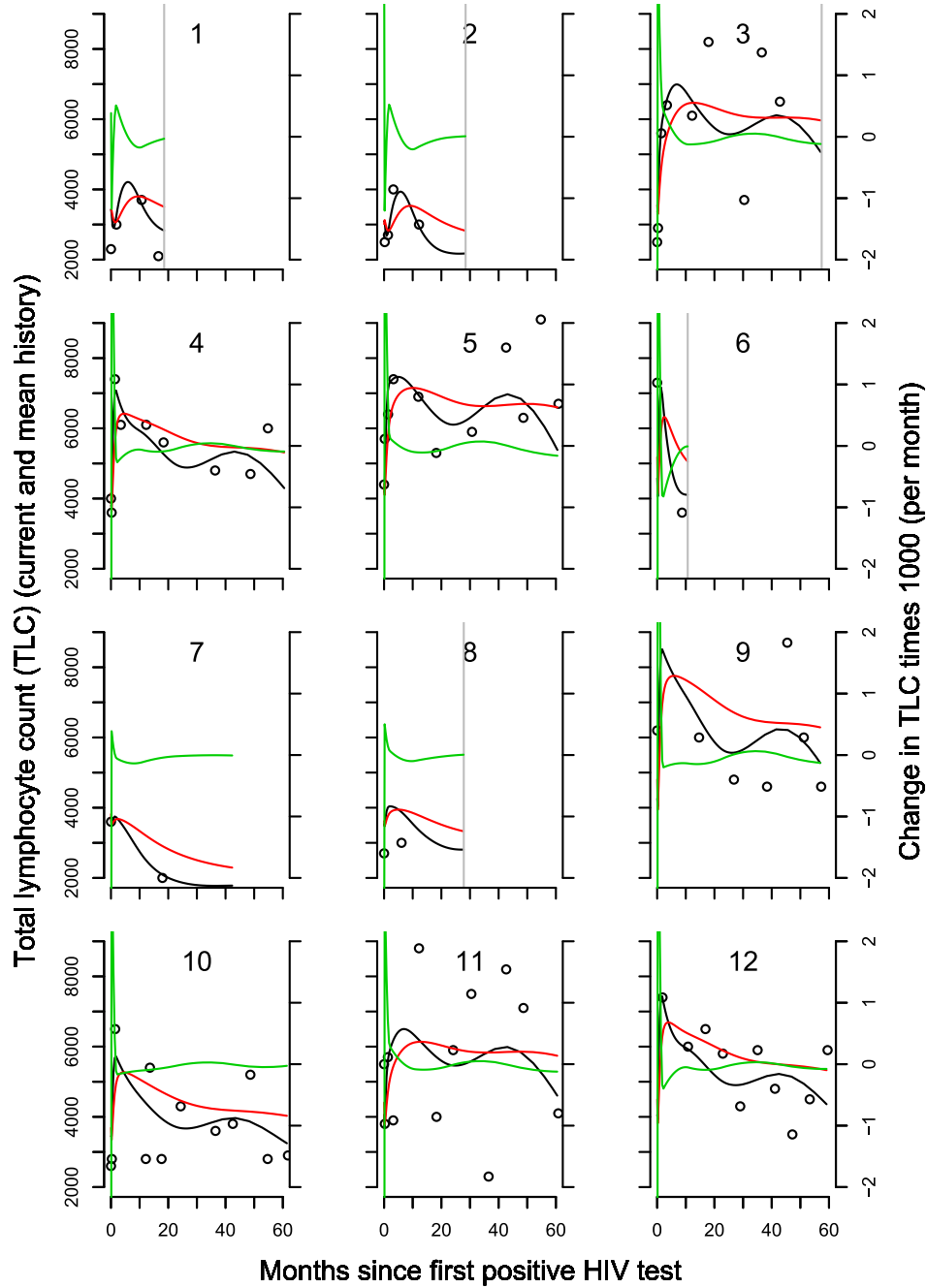

FIG. 3. *Longitudinal profiles of TLC for 16 infants with $q = 6$. Circles mark the observed values. The solid black line indicates the fitted trajectory from the model. The red line represents the fitted cumulative value. The green line represents the fitted slope. A vertical grey line indicates time of death if it was observed.*

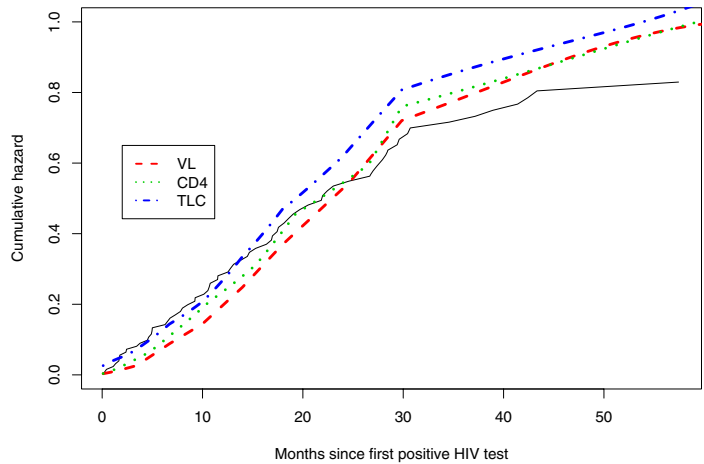

Fig. 4. *Fitted cumulative hazard curves for the selected models for TLC, viral load and CD4 percent. The solid and dashed stepped lines represent the Kaplan–Meier fit with 95% confidence intervals, with the small vertical lines representing censored observation times.*

This model gives interpretable parameters, although these values are difficult to translate into practice. For example, although we estimate that a 10-fold (1 log unit) increase per month in viral load is associated with a 2.8-fold increase in the hazard, we cannot easily measure the instantaneous change in viral load in practice. To do so, we would have to measure it frequently, which is not feasible in practice, especially in resource-poor settings. However, this approach may indicate if there is more information available in collected information than just the current value and what that information might be (change versus average history, for example).

We have proposed a model that provides additional insight into the biological processes of disease progression by linking the hazard of the event to the rate of change or cumulative history of a biomarker. This model expands the possibilities of examining disease progression within the class of joint longitudinal and survival models.

## APPENDIX: SAMPLING FROM THE POSTERIOR

We use Gibbs sampling to sample from the joint posterior distribution of the parameters: $\beta$, $\alpha$, $\gamma$, $\gamma_s$, $\gamma_h$, $\lambda$, $b_o$ and $V_0$. The joint posterior does not have a closed form; however, given that the conditional posteriors either have a closed form or are log-concave, implementation of the Gibbs sampler is straight-forward. Let $D$ denote the data and *rest* denote the remaining

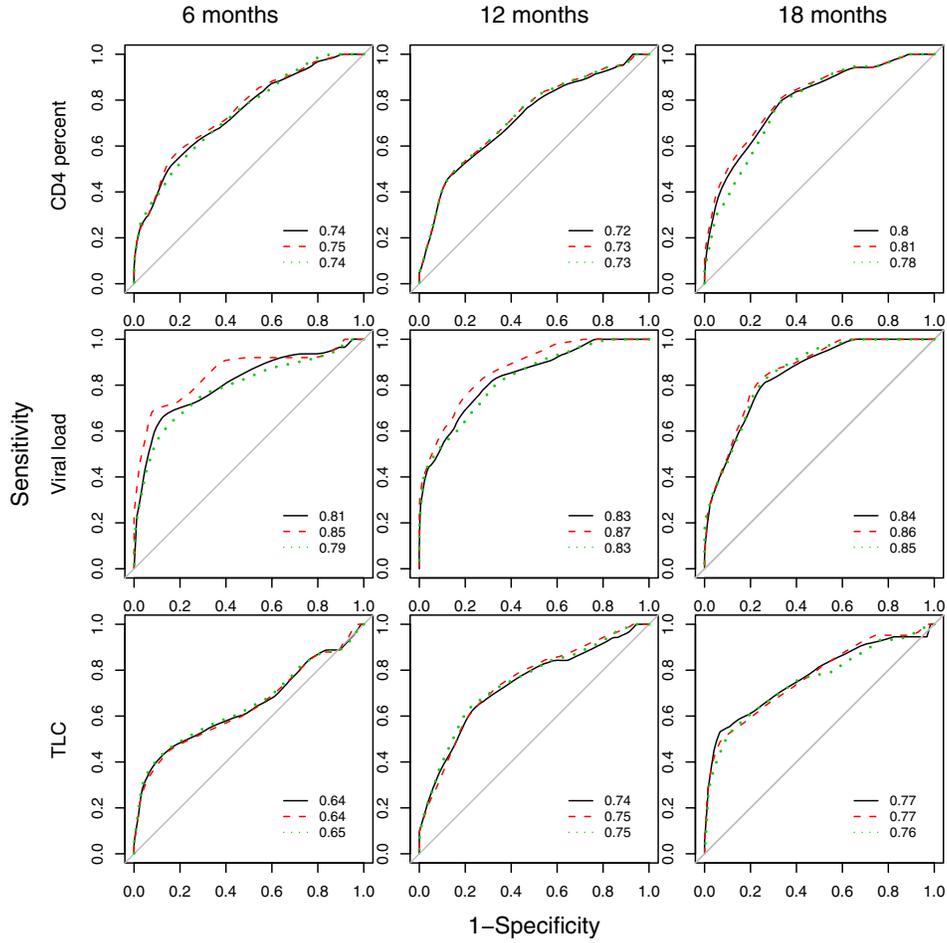

Fig. 5. *ROC curves for predicting death within one year based on the hazard function calculated at 6, 12 and 18 months after initial positive HIV test for current value (solid line), current value and slope (dashed line) and current value and history (dotted line) models. The areas under the ROC curves are shown with each plot.*

parameters. Then at each iteration of the Gibbs sampler, we proceed as follows:

1. Let $\beta_{il} = (\beta_{il1}, \ldots, \beta_{ilq})', i = 1, \ldots, N, l = 1, \ldots, L$. Then use ARS to sample $[\beta_{il}|rest, D]$ from

$$p(\beta_i|rest, D) \propto \exp\left\{-\frac{1}{2}\left[\sum_{j=1}^{m_i}(Y_{ij} - \psi_\beta(t_{ij}))'\Sigma^{-1}(Y_{ij} - \psi_\beta(t_{ij})) \right.\right.$$
$$\left.\left. + \sum_{l=1}^{L}(\beta_{il} - b_{0l} - X_i'\alpha_l)'V_{0l}^{-1}(\beta_{il} - b_{0l} - X_i'\alpha_l)\right]\right\}$$

$$\times \exp\bigg\{\nu_i(\gamma'\psi(s_i) + \gamma'_s\psi'(s_i) + \gamma'_h\psi^{(-1)}(s_i) + z'_i\zeta)$$

$$- e^{z'_i\zeta}\sum_{j=1}^{J} H_{ij}(\beta,\gamma,\gamma_s,\gamma_h,\lambda)\bigg\}.$$

2. Sample

$$[V_{0l}^{-1}|rest, D]$$

$$\sim \text{Wishart}\bigg(\bigg(S_{v_{0l}}^{-1} + \sum_{i=1}^{N}(\beta_{il} - b_{0l} - x'_i\alpha)(\beta_{il} - b_{0l} - x'_i\alpha)'\bigg)^{-1},$$

$$N + \nu_{v_{0l}}\bigg).$$

3. Sample

$$[\Sigma^{-1}|rest, D]$$

$$\sim \text{Wishart}\bigg(\bigg(S_\Sigma^{-1} + \sum_{i=1}^{N}\sum_{j=1}^{m_i}(Y_{ij} - \psi(t_{ij}))(Y_{ij} - \psi(t_{ij}))'\bigg)^{-1},$$

$$\sum_{i=1}^{N} m_i + \nu_\Sigma\bigg).$$

4. Let $b_{0l} = (b_{0l1},\ldots,b_{0lq})', i = 1,\ldots,N, l = 1,\ldots,L$. Then sample

$$[b_{0l}|rest, D] \sim N_2(\mu_{b_{0l}}, \Sigma_{b_{0l}}),$$

where

$$\mu_{b_{0l}} = \Sigma_{b_{0l}}\bigg(V_{0l}^{-1}\sum_{i=1}^{N}(\beta_{il} - x'_i\alpha) + A_{1l}^{-1}A_{0l}\bigg)$$

and

$$\Sigma_{b_{0l}} = (NV_{0l}^{-1} + A_{1l}^{-1})^{-1}.$$

5. Use ARS to sample $[\gamma,\gamma_s,\gamma_h|rest, D]$ from

$$p(\gamma|rest, D) \propto \exp\bigg\{\sum_{i=1}^{N}\bigg(\nu_i(\gamma'\psi(s_i) + \gamma'_s\psi'(s_i) + \gamma'_h\psi^{(-1)}(s_i) + z'_i\zeta)$$

$$- e^{z'_i\zeta}\sum_{j=1}^{J} H_{ij}(\beta,\gamma,\gamma_s,\gamma_h,\lambda)\bigg)\bigg\}$$

$$\times \exp\bigg\{-\frac{1}{2}((\gamma,\gamma_s,\gamma_h)' - g_0)'g_1^{-1}((\gamma,\gamma_s,\gamma_h)' - g_0)\bigg\}.$$

6. Sample $[\lambda_j|rest, D] = \text{gamma}(d_{0j} + n_j, \sum_{i=1}^{N} e^{z_i'\zeta} H_{ij}(\beta, \gamma, \gamma_s, \gamma_h, 1) + d_{1j})$, where $n_j$ is the number of events in the $j$th interval and $H_{ij}(\beta, \gamma, 1)$ is $H_{ij}(\beta, \gamma, \gamma_s, \gamma_h, \lambda)$ evaluated with $\lambda_j = 1$.

7. Sample

$$[\alpha|rest, D] \sim N_p(\mu_\alpha, \Sigma_\alpha),$$

where

$$\mu_\alpha = \Sigma_\alpha \left( \sum_{i=1}^{N} X_i \mathbf{1}_q V_{0l}^{-1}(\beta_{il} - b_{0l}) + C_1^{-1} C_0 \right)$$

and

$$\Sigma_\alpha = \left( \sum_{i=1}^{N} X_i \mathbf{1}_q V_{0l}^{-1} \mathbf{1}_q X_i' + C_1^{-1} \right)^{-1},$$

where $\mathbf{1}_q$ is a vector of ones of length $q$.

**Acknowledgments.** The author thanks an Associate Editor and the referees for their comments that have greatly improved the presentation of the article.

University of Washington
Seattle, Washington 98195
USA
E-mail: elizab@u.washington.edu